\documentclass[aps,onecolumn,pre,showpacs,groupedaddress,showkeys,amsmath,amssymb,superscriptaddress]{revtex4-2}

\usepackage{float}
\usepackage{hyperref}
\usepackage{graphicx}
\usepackage{epstopdf}
\usepackage[caption=false]{subfig}
\usepackage{amsmath}
\usepackage{xcolor}
\usepackage{soul}
\usepackage{upgreek}

\begin{document}

\title{Numerical method to characterise capsule membrane permeability \\ for controlled drug delivery}

\author{Clément Bielinski}
\affiliation{Biomechanics and Bioengineering Laboratory,\\
Université de Technologie de Compiègne, CNRS,\\
60200 Compiègne, France}
\author{Badr Kaoui}
\email{badr.kaoui@utc.fr}
\affiliation{Biomechanics and Bioengineering Laboratory,\\
Université de Technologie de Compiègne, CNRS,\\
60200 Compiègne, France}

\date{\today}

\begin{abstract}
Design and characterisation of capsules is not an easy task owing to the multiple involved preparation factors and parameters.
Here, a novel method to characterise capsule membrane permeability to solute molecules by an inverse approach is proposed.
Transport of chemical species between the capsule core and the surrounding medium through the membrane is described by the Fick's second law with a position-dependent diffusion coefficient.
Solutions are computed in spherical coordinates using a finite difference scheme developed for diffusion in multilayer configuration.
They are validated using semi-analytical solutions and fully three-dimensional lattice Boltzmann simulations.
As a proof of concept, the method is applied to experimental data available in the literature on the kinetics of glucose release and absorption to determine the membrane permeability of capsules.
The proposed method is easy to use and determines correctly the permeability of capsule membranes for controlled drug release and absorption applications.
\end{abstract}


\maketitle

\section{Introduction}
\label{sec:problem_statement}
\begin{figure}[b]
    \centering
    \includegraphics*[width=0.4\textwidth]{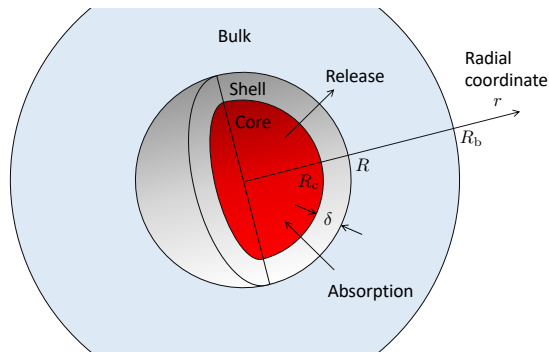}
    \caption{Scheme showing the structure of a core-shell capsule with radius $R$. $R_{\rm c}$ is the radius of the core and $R_{\rm b}$ the radius of the bulk where the capsule is immersed. The shell is supposed to be homogeneous with a uniform thickness $\delta$. The scheme also shows the mass transport directions of the release and absorption of a solute.}
    \label{fig:scheme}
\end{figure}

The last decades have known an increasing interest in using capsules, closed polymeric membranes, to encapsulate active agents, which has led to a large amount of publications and patents.
Capsules are composed of an inner core, where active molecules are encapsulated, and an outer protective thick shell or thin membrane, as illustrated in Fig.~\ref{fig:scheme}.
The main advantage of capsules is their ability to monitor the release of their cargo in a controlled manner.
Moreover, their membrane improves the mechanical robustness and protects the encapsulated material against undesired external chemical and mechanical damages.
Applications of capsules span from food industry to medicine, with capsule size ranging from few millimeters down to few nanometers.
They are fabricated through diverse techniques \cite{Ghosh2006,Sagis2015}, including emulsification in flow focusing microfluidic devices \cite{Chu2013,Galle2016,Li2018}.
Capsules with more complex structures are being developed by adding multiple extra layers or chemically reinforced shells to meet specific performance requirements \cite{Daehne2001,Qiu2001,Zarket2017}.
Capsules are also used as miniaturised bioreactors, where living cells are enclosed to carry biochemical reactions as a response to specific applied stimuli \cite{Ma2013,Fischer2020}.

Design and optimisation of capsules is challenging due to the multiple physico-chemical and geometrical parameters involved.
The release kinetics of the encapsulated solute is influenced by the capsule membrane properties, which have to be optimised in order to deliver molecules at desired rates.
For example, for medical applications, a poorly designed capsule may lead to an inefficient treatment, and even worse, to drug concentration exceeding the toxicity threshold.
The main role of capsule membrane is to moderate the exchange of chemical species between the capsule inner core and the outer surrounding medium.
This depends strongly on the permeability of the membrane, which is paradoxically not known a priori.
It is dictated by multiple factors, such as the local curvature, the porosity and the tortuosity of the membrane resulting from the quality of the cross-linking of the polymer, of which the membrane is made.
Thus, the permeability is not a universal determined property.
It depends rather on the material and the conditions under which a capsule is fabricated.
It differs from a population of capsules to another due to slight accidental changes in the fabrication process.
The capsule permeability cannot be measured directly because of the impossibility of accessing the protected inner core, for example, to place a probe apparatus.
The closed topological character of the capsules leaves only the option to access physically their external surrounding environment.

The state-of-the-art approach to determine capsule permeability consists in measuring the evolution over time of the solute concentration in the external surrounding medium, which is then fit with an equation.
This approach has been employed in many studies \cite{Dembczynski2000,Koyama2004,Yao2006,Ameloot2011,Henning2012a,Leick2011,Shahravan2012,Rolland2014,BenMessaoud2016}.
For the release case, the most widely used equation is the analytical solution of the Fick's second law for solute diffusion from a sphere to a well-stirred solution of finite volume \cite{Crank1975},
\begin{equation}
\frac{C_t}{C_{\rm eq}} = 1 - \sum_{n=1}^{\infty} \frac{6\alpha(1+\alpha)}{9+9\alpha+q_n^2\alpha^2}\exp\left( -\frac{Dq_n^2}{R^2}t\right)\text{,}
\label{eq:analytical_well_stirred_release}
\end{equation}
where $C_t$ is the solute concentration in the bulk at time $t$, which is assumed to be uniform due to stirring.
$C_{\rm eq}$ is the expected concentration at equilibrium, \textit{i.e.} as $t\rightarrow \infty$.
$R$ is the radius of the sphere, $D$ is the solute diffusion coefficient in the sphere, and $\alpha = {V_{\rm b}}/{V}$ is the ratio of the volumes of the bulk and the sphere, when assuming the partition coefficient is unity.
$q_n$ is the $n^{\rm th}$ non-zero positive root of $\tan q_n = 3q_n/(3 + \alpha q_n^2)$.
$C_{\rm eq}$ depends on the initial concentration $C_0$ within the sphere.
It can be deduced from the mass conservation $C_0V = C_{\rm eq}(V_{\rm b}+V)$ that leads to $C_{\rm eq} = C_0/(\alpha + 1)$.
For the absorption case, namely for diffusion from a well-stirred bulk solution to the capsule, the analytical solution is rather \cite{Crank1975},
\begin{equation}
\frac{C_t}{C_\mathrm{0}}=\frac{\alpha}{1+\alpha}\left[1+\sum_{n=1}^\infty \frac{6\left(1+\alpha\right)}{9+9\alpha+q_n^2\alpha^2}\exp\left(-\frac{D q_n^2t}{R^2}\right)\right]\text{,}
\label{eq:analytical_well_stirred_absorption}
\end{equation}
where $C_0$ is, here, the initial concentration in the bulk.
However, Eqs.~\eqref{eq:analytical_well_stirred_release} and \eqref{eq:analytical_well_stirred_absorption} are not derived originally for a capsule with a composite core-shell structure, but simply for a homogeneous sphere (see Ref.~\cite{Bielinski2021b} for the mathematical derivation).
The diffusion coefficient $D$ is then considered by most authors as the effective diffusion coefficient of the overall capsule (core and membrane altogether).
In this way, the fit enables the determination of an effective diffusion coefficient for the whole capsule, and not specifically for the membrane.
However, the membrane permeability is a key parameter in designing, for example, capsules as bioreactors encapsulating cells and moderating the nutrients exchange rate via their membrane \cite{Ma2013}.
Only few authors have attempted to fit their experimental data with mathematical models derived for membranes.
Kondo \cite{Kondo1990} has used a model for a planar membrane by assuming negligible effects of the capsule curvature when the aspect ratio of the membrane thickness $\delta$ to the capsule radius $R$ tends to zero ($\delta/R\rightarrow0$).
Henning \textit{et al.} \cite{Henning2012b} have measured concentration profiles using the NMR technique, and have fitted their experimental data with numerical solutions computed with the Matlab \textit{pdepe} solver \cite{Matlab2007}, instead of Eq.~\eqref{eq:analytical_well_stirred_absorption}.

The present article proposes a novel straightforward and correct methodology to characterise capsules permeability by an inverse approach, using solutions of the diffusion equation computed with the finite difference method.
This article is organised as follows.
The proposed method is explained in details in Sec.~\ref{sec:method}, where the governing equations, the numerical scheme and its validation are presented.
The efficiency of the suggested method is then demonstrated in Sec.~\ref{sec:results} by applying it to numerical data and available experimental data in the literature (\textit{e.g.} Refs.~\cite{Koyama2004,Rolland2014}).
Finally, conclusions are drawn in Sec.~\ref{sec:conclusions}.
\section{Method}
\label{sec:method}
\textit{Governing equations} - Diffusion of chemical species in a medium composed of an inner core and an outer domain separated by a shell that moderates mass exchange between these two compartments is governed by the Fick's second law in absence of advection,
\begin{equation}
\frac{\partial c}{\partial t} = \nabla \cdot \left( D\nabla c\right)\text{,}
\label{eq:diffusion_equation}
\end{equation}
where $t$ is the time and $c$ the local concentration.
$D$ is the position-dependent diffusion coefficient taking different values in the core ($0 \leq r \leq R_\mathrm{c}$), in the shell ($R_\mathrm{c} < r \leq R$) and in the surrounding bulk ($R < r \leq R_\mathrm{b}$),
\begin{equation}
D(r)=\begin{cases}
D_\mathrm{c}, & 0 \leq r \leq R_\mathrm{c}\\
D_\mathrm{m}, & R_\mathrm{c} < r \leq R\\
D_\mathrm{b}, & R < r \leq R_\mathrm{b}\\
\end{cases}
,
\end{equation}
with $r$ being the distance from the centre of the capsule.
As the shell is a cross-linked gel or a porous rigid wall, its diffusivity to the solute $D_\mathrm{m}$ is expected to be lower than the diffusivities in the core and the bulk, if these are liquids.
We refer to the diffusion coefficient $D_{\rm m}$ as the permeability of the capsule membrane to solute, see for example \cite{Kondo1990,Henning2012b}, while having in mind other definitions such as the ratio of the diffusion coefficient $D_{\rm m}$ to the shell thickness $\delta$ \cite{Bielinski2021a}. All these definitions are adapted to cases where the shell thickness is not negligible compared to the size of the capsule $\delta / R \gg 0.1$.
The problem presents a spherical symmetry.
Thus, the diffusion is expected to be purely radial with $c = c(r,t)$.
Equation \eqref{eq:diffusion_equation} is consequently expressed only as a function of its radial derivative terms,
\begin{equation}
\frac{\partial c}{\partial t} = \frac{1}{r^2}\frac{\partial}{\partial r}\left(r^2D\frac{\partial c}{\partial r}\right) = \frac{2D}{r}\frac{\partial c}{\partial r} + \frac{\partial D}{\partial r}\frac{\partial c}{\partial r} + D \frac{\partial^2 c}{\partial r^2}\text{.}
\label{eq:diffusion_equation_axisymmetric}
\end{equation}

\textit{Initial condition} - Diffusion from and to capsules is considered in the present study.
For the release case, \textit{i.e.} diffusion from the capsule to the bulk, the core of the capsule is initially loaded with a uniform concentration $C_0$, and the concentration is zero elsewhere,
\begin{equation}
c(r,t=0)=\begin{cases}
C_0, & 0 \leq r \leq R_\mathrm{c}\\
0, & r > R_\mathrm{c}
\end{cases}
.
\end{equation}
This establishes a concentration gradient that triggers diffusion from the core to the bulk.
For the absorption case, \textit{i.e.} diffusion into the capsule, the core and the membrane are initially free of solute while the concentration is uniform and equal to $C_0$ in the bulk,
\begin{equation}
c(r,t=0)=\begin{cases}
0, & r \leq R\\
C_0, & r > R
\end{cases}
.
\end{equation}
This condition, on the contrary, induces diffusion from the bulk to the capsule.

\textit{Boundary conditions} - The resulting spatial distribution of the concentration depends also on the boundary conditions.
Here, unsteady continuous concentration and mass flux boundary conditions emerge in the absence of interfacial mass resistance at the core-shell interface $r=R_{\rm c}$,
\begin{align}
& c(R_\mathrm{c}^-,t) = c(R_\mathrm{c}^+,t) \quad \text{and} \quad D_\mathrm{c} \left.\frac{\partial c}{\partial r}\right|_{r=R_\mathrm{c}^-} = D_\mathrm{m} \left.\frac{\partial c}{\partial r}\right|_{r=R^+_\mathrm{c}},
\label{eq:continuity_core_membrane}
\end{align}
and at the shell-bulk interface $r=R$,
\begin{align}
& c(R^-,t) = c(R^+,t) \quad \text{and} \quad D_\mathrm{m} \left.\frac{\partial c}{\partial r}\right|_{r=R^-} = D_\mathrm{b} \left.\frac{\partial c}{\partial r}\right|_{r=R^+},
\label{eq:continuity_membrane_bulk}
\end{align}
where the superscripts ($-$) and (${\rm +}$) refer to the inner and outer sides of the interfaces, respectively.
These boundary conditions relax the constraints of imposing perfect sink or well-stirred boundary conditions.
These latter could, in principle, be fulfilled in an experiment, but not necessarily under operational conditions.
They limit the computation to the capsule by excluding the bulk.
The perfect sink conditions can be modeled with a Dirichlet boundary condition at the capsule surface,
\begin{equation}
c(R,t)=0,
\end{equation}
and the well-stirred solution with the boundary condition,
\begin{equation}
\frac{\partial c}{\partial t} = - D_{\rm m}\left( \frac{3R^2}{R_{\rm b}^3 - R^3}\right)  \left.\frac{\partial c}{\partial r}\right|_{r=R},
\label{eq:stirring_bc}
\end{equation}
to be set at the surface of the capsule (see Ref.~\cite{Bielinski2021b} for the mathematical derivation of this equation).
When the study considers three layers (core, shell and bulk), no mass flux is imposed at the domain edge ($r=R_\mathrm{b}$),
\begin{equation}
\left. \frac{\partial c}{\partial r}\right|_{r=R_\mathrm{b}} = 0.
\end{equation}
The condition,
\begin{equation}
\left.\frac{\partial c}{\partial r}\right|_{r=0} = 0,
\end{equation}
is set at the origin ($r=0$) in all the simulations to fulfill the radial symmetry.

\begin{figure}[H]
    \centering
    \includegraphics*[width=0.48\textwidth]{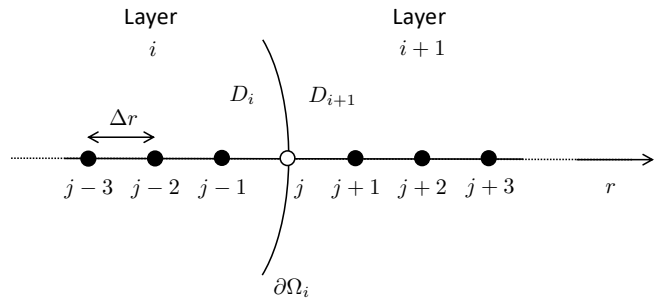}
    \caption{Scheme showing the spatial discretisation in the vicinity of the interface $\partial \Omega_i$ between two adjacent layers $i$ and $i+1$, where the solute diffusivities are respectively $D_{i}$ and $D_{i+1}$. The inner points are represented with filled circles, and the point at the interface is shown as an empty circle. The space step $\Delta r$ is uniform in the whole domain.}
    \label{fig:discretisation}
\end{figure}

\textit{Numerical FD solver} - The governing equations are solved numerically using the finite difference method (FD), while considering continuity of both the concentration and the mass flux at the inner and the outer interfaces of the shell.
The discretisation of the equations is adopted from the scheme proposed by Hickson \textit{et al.} \cite{Hickson2011} for diffusion in one-dimensional multilayer slabs.
It is adapted, here, to one-dimensional spherical coordinates that introduce additional terms.
The diffusion equation in spherical coordinates, Eq.~\eqref{eq:diffusion_equation_axisymmetric}, is discretised at the inner points of each subdomain (represented as filled circles in Fig.~\ref{fig:discretisation}) by a central finite difference approximation in space.

The capsule is considered to be constituted of homogeneous layers, \textit{i.e.} layers of constant diffusivity, which leads to,
\begin{equation}
\frac{\partial c_j}{\partial t} = \frac{D_i}{\Delta r^2} \left[\left(\frac{j-1}{j}\right)c_{j-1}  -2c_j + \left(\frac{j+1}{j}\right)c_{j+1}\right]\text{,}
\label{eq:diffusion_equation_FD_scheme}
\end{equation}
for the discretisation at the inner points of layer $i$.
Here, $c_j$ is the local concentration at a given inner point $j$ of the layer $i$, $D_i$ is the diffusion coefficient in the layer $i$, and $\Delta r$ the space step, which is set uniform in the whole domain.
The discretisation of the domain proposed by Hickson \textit{et al.} assumes one grid-point $j$ lies on the interface $\partial \Omega_i$ between the layers $i$ and $i+1$, as illustrated in Fig.~\ref{fig:discretisation}.
Taking a central difference scheme in space to discretise Eq.~\eqref{eq:diffusion_equation_axisymmetric} at the interface $\partial \Omega_i$, and then taking first-order forward and backward differences for the spatial derivatives of the concentration gives,
\begin{equation}
\frac{\partial c_j}{\partial t} = \frac{1}{2\Delta r^2} \left\{ D_i\left(\frac{j-1}{j}\right)^2 c_{j-1} - \left[D_{i}\left(\frac{j-1}{j}\right)^2 + D_{i+1}\left(\frac{j+1}{j}\right)^2 \right]c_j + D_{i+1}\left(\frac{j+1}{j}\right)^2c_{j+1} \right\} \text{.}
\label{eq:conditions_interface}
\end{equation}
Equations \eqref{eq:diffusion_equation_FD_scheme} and \eqref{eq:conditions_interface} can be written together under the form of a single linear system,
\begin{align}
    \frac{\mathrm{d}}{\mathrm{d}t}
        \begin{bmatrix}
        \vdots\\
        c_{j-3}\\
        c_{j-2}\\
        c_{j-1}\\
        c_j\\
        c_{j+1}\\
        c_{j+2}\\
        c_{j+3}\\
        \vdots
        \end{bmatrix}
     =
    \begin{bmatrix}
        \ddots &  &  &  &  &  &  &  & \\
        & \ddots &  &  &  &  &  &  &\\
        \cdots & \frac{j-3}{j-2}\chi_i & -2\chi_i & \frac{j-1}{j-2}\chi_i & 0 & 0 & 0 & 0 & \cdots\\
        \cdots & 0 & \frac{j-2}{j-1}\chi_i & -2\chi_i & \frac{j}{j-1}\chi_i & 0 & 0 & 0 &\cdots\\
        \cdots & 0 & 0 & a & -(a + b) & b & 0 & 0 &\cdots\\
        \cdots & 0 & 0 & 0 & \frac{j}{j+1}\chi_{i+1} & -2\chi_{i+1} & \frac{j+2}{j+1}\chi_{i+1} & 0 &\cdots\\
        \cdots & 0 & 0 & 0 & 0 & \frac{j+1}{j+2}\chi_{i+1} & -2\chi_{i+1} & \frac{j+3}{j+2}\chi_{i+1} &\cdots\\
        &  &  &  &  &  &  & \ddots &  \\
        &  &  &  &  &  &  &  & \ddots\\
        \end{bmatrix}
    \begin{bmatrix}
        \vdots\\
        c_{j-3}\\
        c_{j-2}\\
        c_{j-1}\\
        c_j\\
        c_{j+1}\\
        c_{j+2}\\
        c_{j+3}\\
        \vdots
        \end{bmatrix}
        ,
    \label{eq:linear_system}
\end{align}
where $c_j$ is the concentration at the interface $\partial \Omega_i$ located at the node $j$ (see Fig.~\ref{fig:discretisation}), $\chi_i = \frac{D_i}{\Delta r^2}$, $\chi_{i+1} = \frac{D_{i+1}}{\Delta r^2}$, $a = \frac{\chi_i}{2}\left(\frac{j-1}{j}\right)^2$, and $b = \frac{\chi_{i+1}}{2}\left(\frac{j+1}{j}\right)^2$.
The concentration at iteration $n+1$ is computed using a standard Euler scheme,
\begin{equation}
{\rm C}^{n+1} = {\rm C}^n + \Delta t {\rm M C}^n\text{,}
\label{eq:linear_system_compact}
\end{equation}
where ${\rm C}^n$ is the column vector containing the values of the concentration at each grid-point at iteration $n$, $\Delta t$ is the timestep, and ${\rm M}$ the tridiagonal matrix in Eq.~\eqref{eq:linear_system}.
Here, ${\rm M}$ includes extra terms compared to the matrix given in Ref.~\cite{Hickson2011} because of the spherical coordinates for which is derived.

\begin{figure}[b]
    \centering
    \subfloat[\label{subfig:validation_semi_analytic}]{\includegraphics[width=0.45\textwidth]{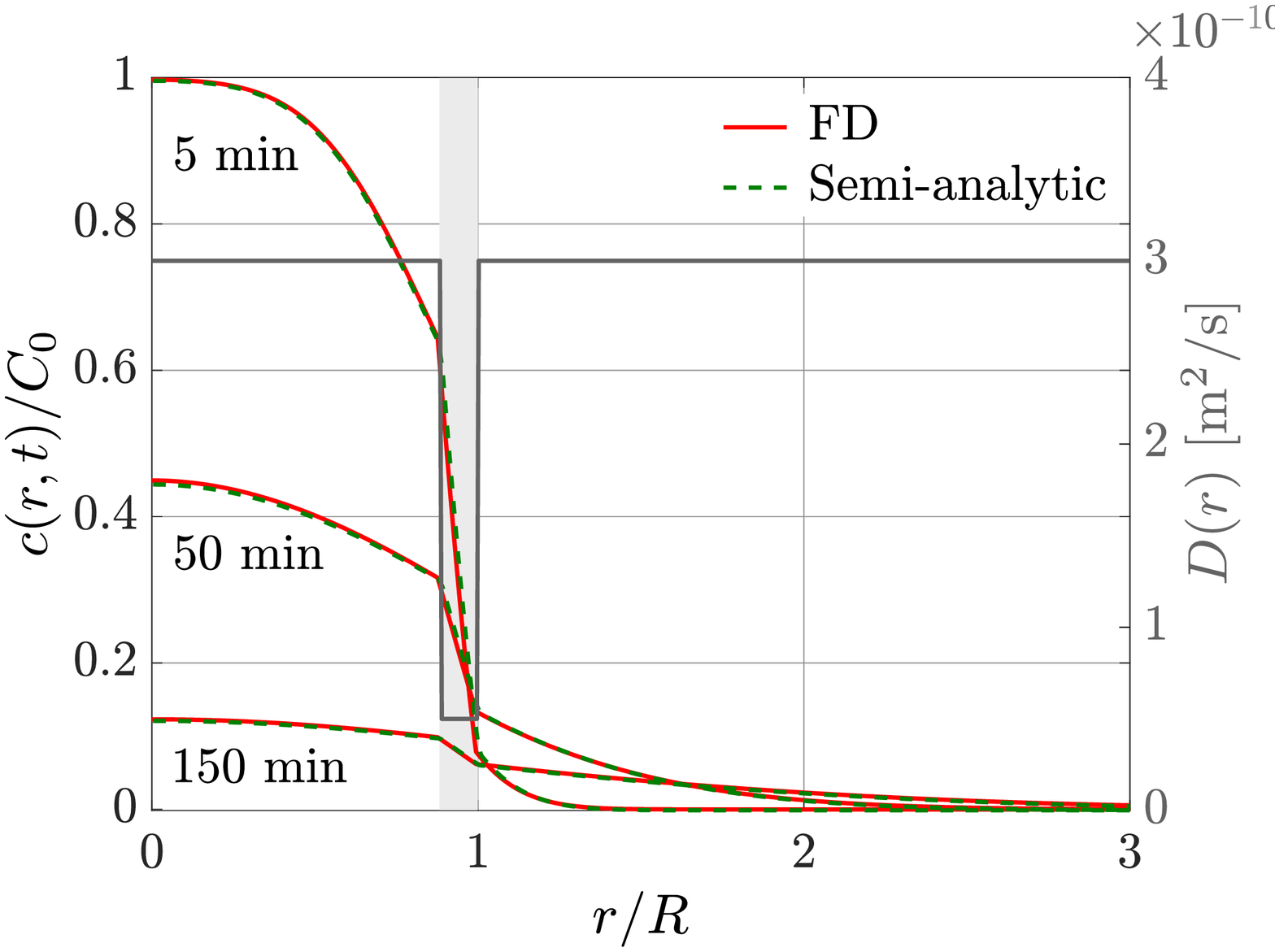}}
    \quad\quad
    \subfloat[\label{subfig:mesh_dependence}]{\includegraphics[width=0.45\textwidth]{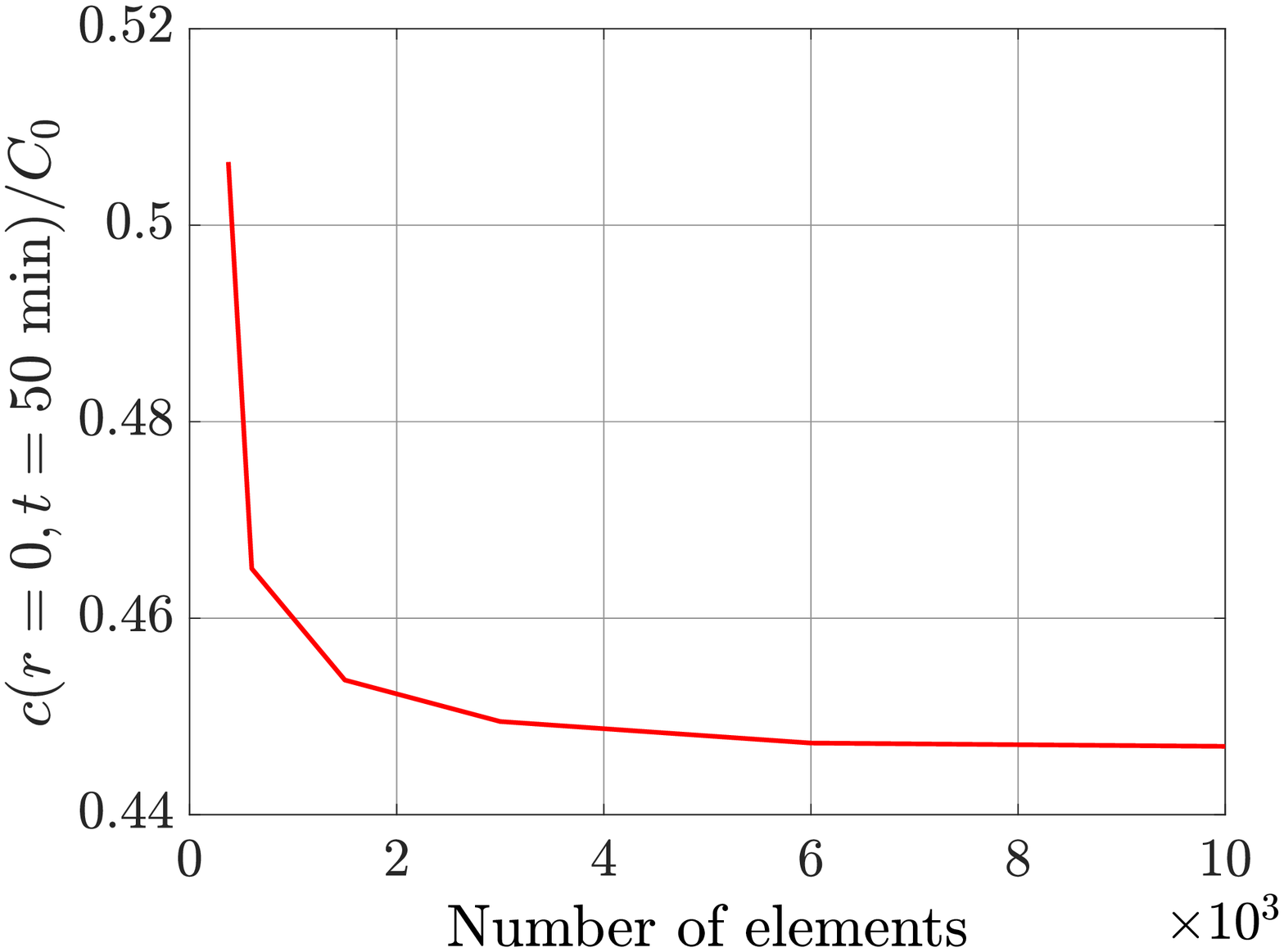}}
    \caption{(a) Concentration profiles computed at times $t = 5, 50$ and $150\,{\rm min}$. The red solid line curves represent the numerical solution of Eq.~\eqref{eq:diffusion_equation_axisymmetric} obtained with the proposed finite difference scheme (FD), while the green dashed lines represent the semi-analytical solution taken from Ref.~\cite{Kaoui2018}. The dark grey solid line gives the local solute diffusion coefficient $D(r)$. The light grey rectangle depicts the location of the membrane. Excellent agreement is achieved between the FD and the semi-analytical solutions. (b) Concentration computed at the center of the capsule at $t = 50$\,min for various grid resolutions. The result converges to a unique value at high grid resolutions.}
    \label{fig:validation}
\end{figure}
\textit{Validation} - The FD scheme is validated by comparing the computed concentration profiles with semi-analytical solutions of Ref.~\cite{Kaoui2018}, with $R_\mathrm{c} = 1.5\,{\rm mm}$, $\delta = 0.2\,{\rm mm}$, $R_\mathrm{b} = 30\,{\rm mm}$, $D_\mathrm{m} = 0.5\times10^{-10}\,{\rm m}^2/{\rm s}$, and $D_\mathrm{c} = D_\mathrm{b} = 3\times10^{-10}\,{\rm m}^2/{\rm s}$.
The concentration profiles at times $t = 5, 50$ and $150\,{\rm min}$ are presented in Fig.~\ref{subfig:validation_semi_analytic}.
For clarity reasons, the figure is truncated at $r/R = 3$ beyond which the concentration is almost zero.
The membrane location is depicted by the light grey rectangle.
The expected physics is well recovered with continuous and discontinuous variations of respectively the concentration and its derivative across the membrane.
Perfect agreement with the semi-analytical solution is achieved, as well.
The profiles are computed using the Matlab script \texttt{diffusion\_FD.m} provided in Ref.~\cite{Bielinski2021c}.
For this simulation, the space step is set to $\Delta r = 10^{-5}$\,m which corresponds to 3000 elements.
A grid independence study has been conducted to make sure the obtained results are reliable and invariant of grid resolution.
Figure \ref{subfig:mesh_dependence} shows the concentration at the capsule center ($r = 0$) at time $t=50$\,min computed for different grid resolutions using the same simulation parameters as in Fig.~\ref{subfig:validation_semi_analytic}.
The results are not affected when refining the discretization further than $\Delta r = 10^{-5}$\,m (3000 elements), which is a good compromise between accuracy and computing time.

\textit{Permeability characterisation} - The proposed method is based on an inverse analysis approach.
The required inputs are the experimental data providing either the release or the uptake kinetics of a solute, and the geometrical characteristics of the studied capsule.
A wide interval of the fitting parameter space $(D_\mathrm{m}, D_\mathrm{c})$ is scanned, and the resulting output release or uptake curves are then compared against the experimental data to quantify the root mean square error (RMSE),
\begin{equation}
\mathrm{RMSE} = \sqrt{\frac{1}{N}\sum_{i=1}^N\left(C_{\rm exp}(i)-C_{\rm num}(i)\right)^2}\text{,}
\label{eq:rmse}
\end{equation}
where $N$ is the number of the input experimental data points ${C}_{\rm exp}$, and $C_{\rm num}$ the numerical solution.
Only cases with $D_\mathrm{m} \leq D_\mathrm{c}$ are considered because diffusion in the membrane is expected to be slower than in the core, since the role of the membrane is to moderate mass exchange.
The optimal values of the diffusivities $(D_\mathrm{m}, D_\mathrm{c})$ are those leading to the minimal error between the numerical solution and the experimental input data.
They are computed using the Matlab script \texttt{compute\_permeability.m} provided in Ref.~\cite{Bielinski2021d}.
\section{Results}
\label{sec:results}
\subsection{Application based on numerical data}
\label{subsec:application_numerical}
\begin{figure}[t]
    \centering
    \subfloat[\label{subfig:concentration_3D}]{\includegraphics[width=0.85\textwidth]{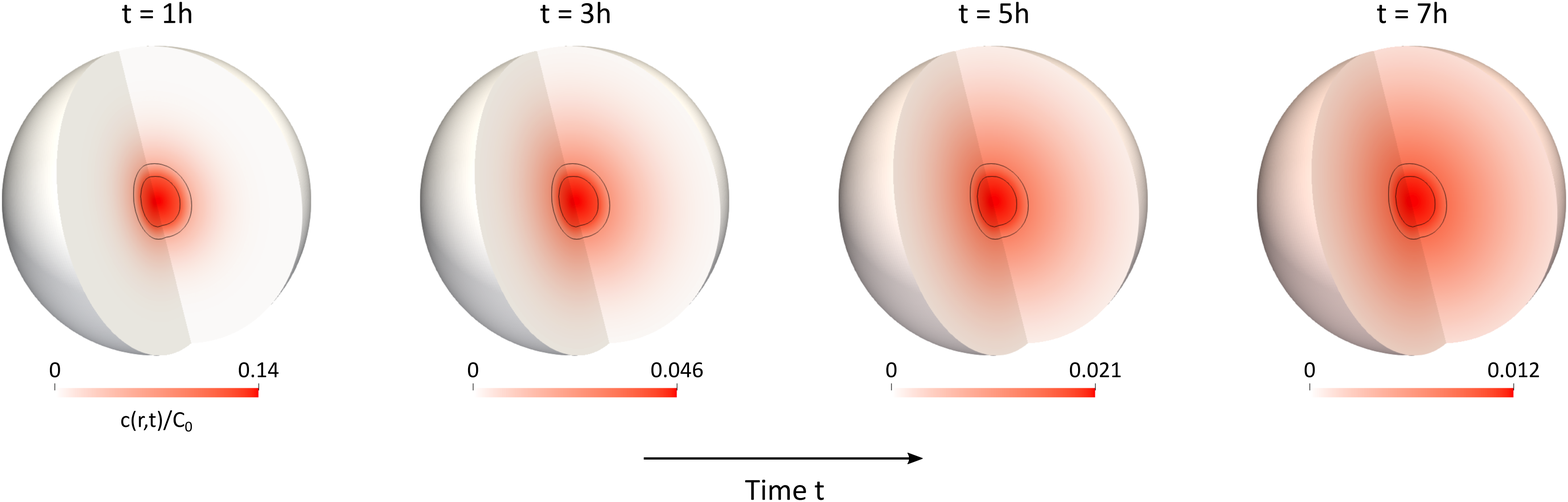}}\\
    \subfloat[\label{subfig:LB_FD_release}]{\includegraphics[width=0.45\textwidth]{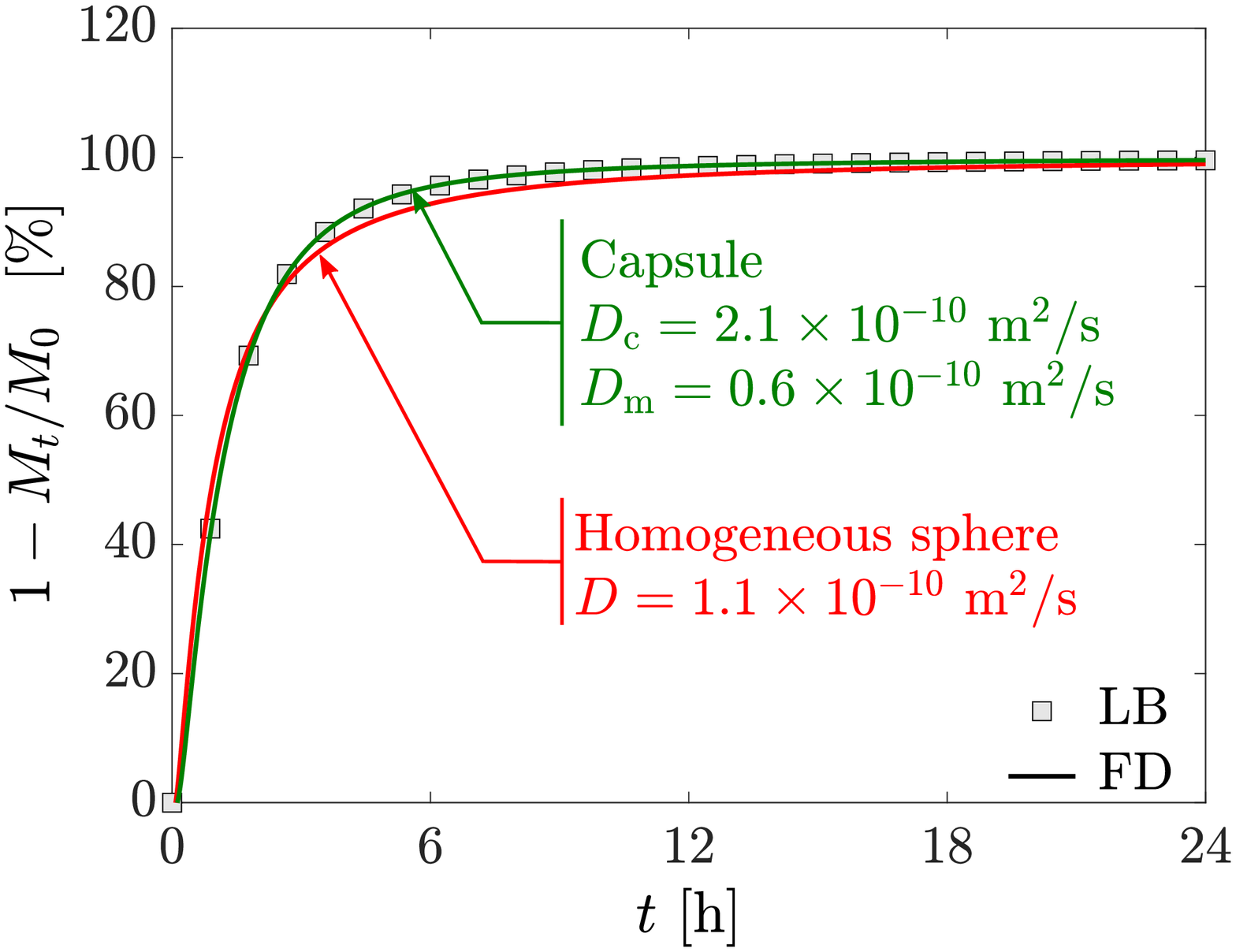}}
    \subfloat[\label{subfig:LB_rmse}]{\includegraphics[width=0.45\textwidth]{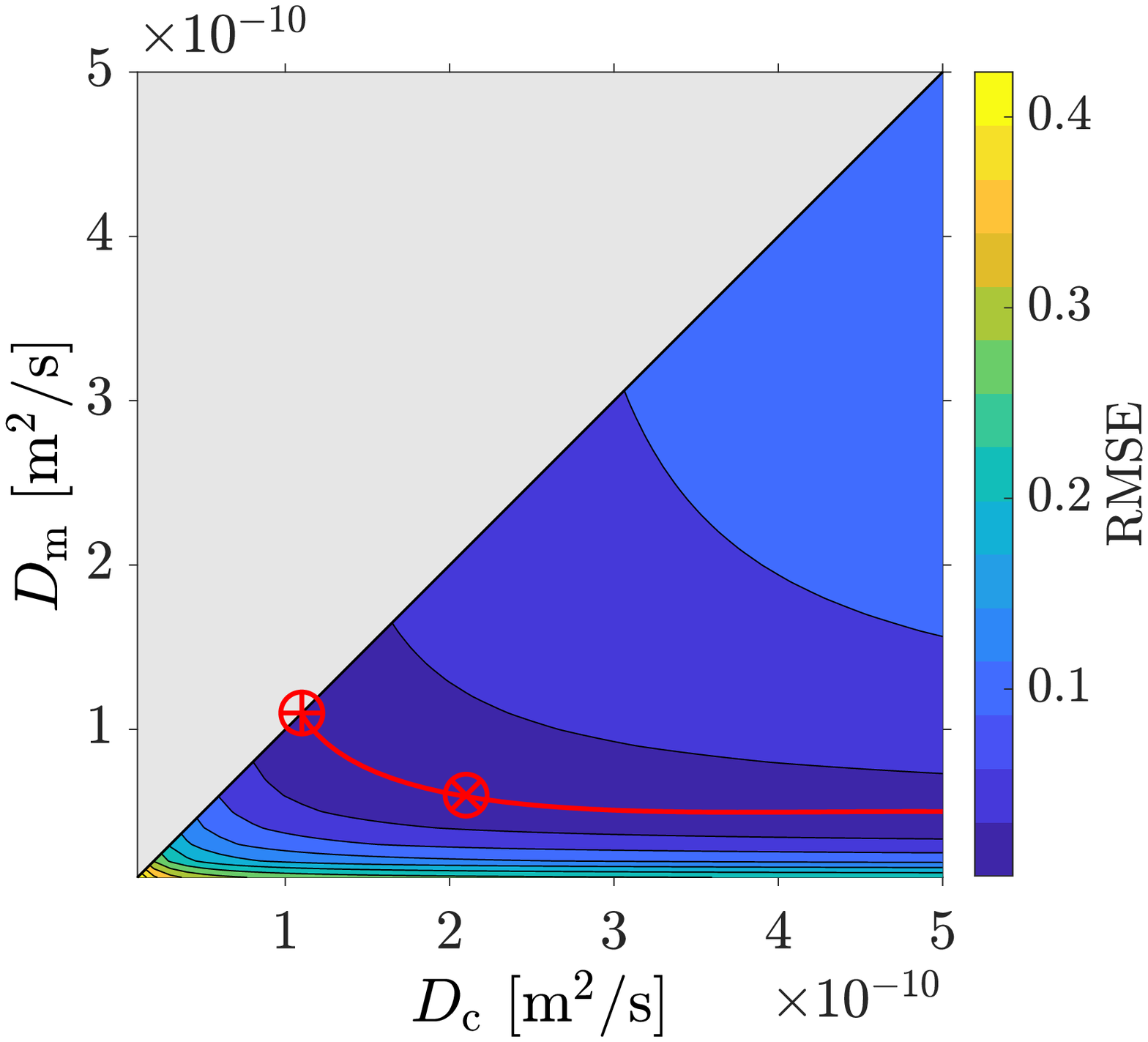}}
    \caption{(a) Snapshots of the concentration field in three dimensions computed at various times by the lattice Boltzmann method (LB) in the case of insulin release from a capsule. The black solid lines delimit the core and the shell of the capsule. The domain is truncated at $r/R_\mathrm{b} = 1/2$ for clarity reasons. The indicated times are rounded down to the nearest hour. (b) The release kinetics curve. Squares correspond to the release kinetics computed by the LB method and used as input data for the characterisation. The green solid curve corresponds to the best fit of the LB data when considering the capsule with a composite structure, and the red solid curve corresponds to the best fit when assuming the capsule as a homogeneous sphere. (c) The RMSE between the LB and the FD solutions over a wide range of the fitting parameters $D_\mathrm{c}$ and $D_\mathrm{m}$. The red solid line gives the value of the membrane diffusivity $D_\mathrm{m}$ that minimises the RMSE at a given core diffusivity $D_\mathrm{c}$. The absolute minimum is obtained at $D_\mathrm{c}=2.1\times10^{-10}\,{\rm m}^2/{\rm s}$ and $D_\mathrm{m}=0.6\times10^{-10}\,{\rm m}^2/{\rm s}$ (symbol $\otimes$), and that match the values set in the LB simulation. These fit better the data than the solution obtained when considering the capsule as a homogeneous sphere with $D=D_\mathrm{m}=D_\mathrm{c}=1.1\times 10^{-10}\,{\rm m}^2/{\rm s}$ (symbol $\oplus$). The core radius, the membrane thickness and the bulk radius are $R_\mathrm{c} = 1\,{\rm mm}$, $\delta = 0.5\,{\rm mm}$ and $R_\mathrm{b} = 12\,{\rm mm}$, respectively.}
    \label{fig:application_LB}
\end{figure}

First, the method is tested on numerical data computed with a fully three-dimensional lattice Boltzmann simulation (LB), for which the diffusion coefficients are set and known in advance by the authors.
Further details on the LB method can be found in Refs.~\cite{Sukop2006,Kruger2016}.
The geometrical parameters used in the LB simulation are $R_\mathrm{c} = 1\,{\rm mm}$, $\delta = 0.5\,{\rm mm}$, and $R_\mathrm{b} = 12\,{\rm mm}$.
The diffusion coefficients are $D_\mathrm{c} = D_\mathrm{b} = 2\times10^{-10}\,\text{m}^2/{\rm s}$ and $D_\mathrm{m} = 0.6\times10^{-10}\,\text{m}^2/{\rm s}$ corresponding to the diffusion of insulin in an aqueous solution and a polymeric membrane, respectively \cite{Morvan1989}.
Continuous boundary conditions, Eqs.~\eqref{eq:continuity_core_membrane} and \eqref{eq:continuity_membrane_bulk}, are considered.
The resulting concentration field in the capsule and in its surrounding medium is given in Fig.~\ref{subfig:concentration_3D} at various times.
The release kinetics is quantified by $\left(1-\frac{M_t}{M_0}\right)$, where $M_t = 4\pi\int_{r=0}^{R}r^2c(r,t)\mathrm{d}r$ is the mass of the solute that remains within the capsule at time $t$.
It is denoted by grey-colored square symbols in Fig.~\ref{subfig:LB_FD_release}.
The characterisation method consists in taking these data (as if they were experimental data ${C}_{\rm exp}$) and fit them with FD numerical solutions using the solute diffusivity in both the membrane and the core $(D_{\rm m},D_{\rm c})$ as fitting parameters, while holding constant the values of the geometrical parameters $R_\mathrm{c}$, $\delta$ and $R_\mathrm{b}$.
Figure \ref{subfig:LB_rmse} reports the RMSE measuring how much the obtained FD solutions deviate from the LB solution.
The red solid line gives the value of the membrane diffusivity $D_\mathrm{m}$ which minimises the RMSE at a given core diffusivity $D_\mathrm{c}$.
The global minimum of the RMSE is obtained at $D_\mathrm{c} = 2.1\times10^{-10}\,{\rm m}^2/{\rm s}$ and $D_\mathrm{m} = 0.6\times10^{-10}\,{\rm m}^2/{\rm s}$ (symbol $\otimes$).
This estimated value of the membrane diffusivity is exactly equal to the value set in the LB simulation, and the small difference obtained for the core diffusivity represents a relative error of only $5\%$, which is satisfactory.
The line $D_\mathrm{m} = D_\mathrm{c}$ corresponds to the case of having a homogeneous sphere, as usually assumed in the literature.
The local minimum of RMSE on this line gives the effective diffusivity $D$ for the whole capsule.
It is located at $D_\mathrm{m} = D_\mathrm{c} = D = 1.1\times10^{-10}\,{\rm m}^2/{\rm s}$ (symbol $\oplus$).
The RMSE is strongly sensitive to changes in the membrane diffusivity $D_\mathrm{m}$, and it is weakly affected by variations of the core diffusivity $D_\mathrm{c}$, especially at low membrane diffusivities.
This means that the release process is mainly controlled by the membrane permeability.

The release curve computed for $D_\mathrm{c} = 2.1\times10^{-10}\,{\rm m}^2/{\rm s}$ and $D_\mathrm{m} = 0.6\times10^{-10}\,{\rm m}^2/{\rm s}$ fits perfectly the input LB release data (see Fig.~\ref{subfig:LB_FD_release}), in contrast to the one of a homogeneous sphere.
This example has allowed us to validate the proposed method and to demonstrate its efficiency to characterise the membrane permeability of capsules.
\subsection{Application to experimental data of glucose release}
\label{subsec:application_experimental}
\begin{figure}[t]
    \centering
    \subfloat[\label{subfig:Rolland_release}]{\includegraphics[width=0.45\textwidth]{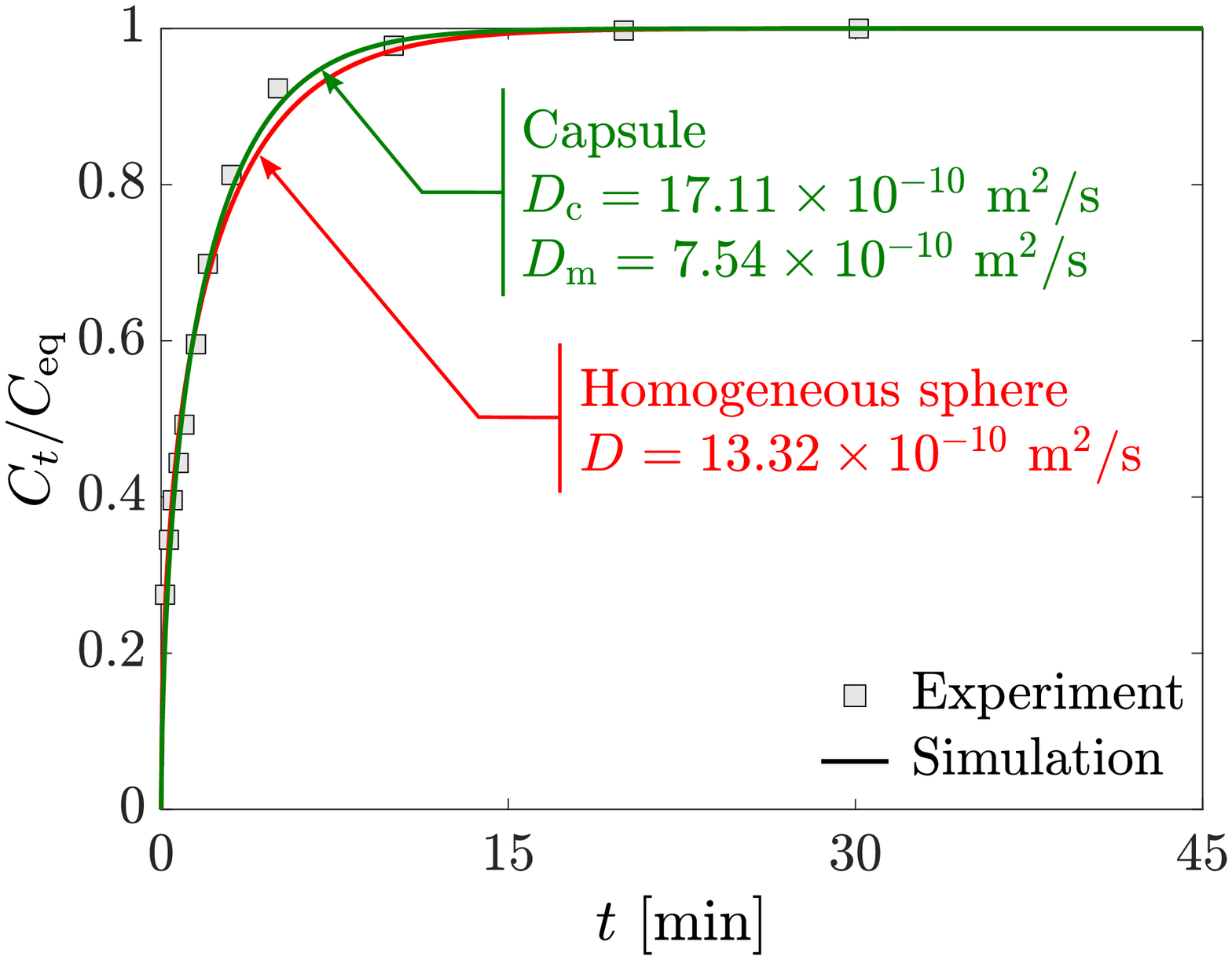}}
    \subfloat[\label{subfig:Rolland_rmse}]{\includegraphics[width=0.45\textwidth]{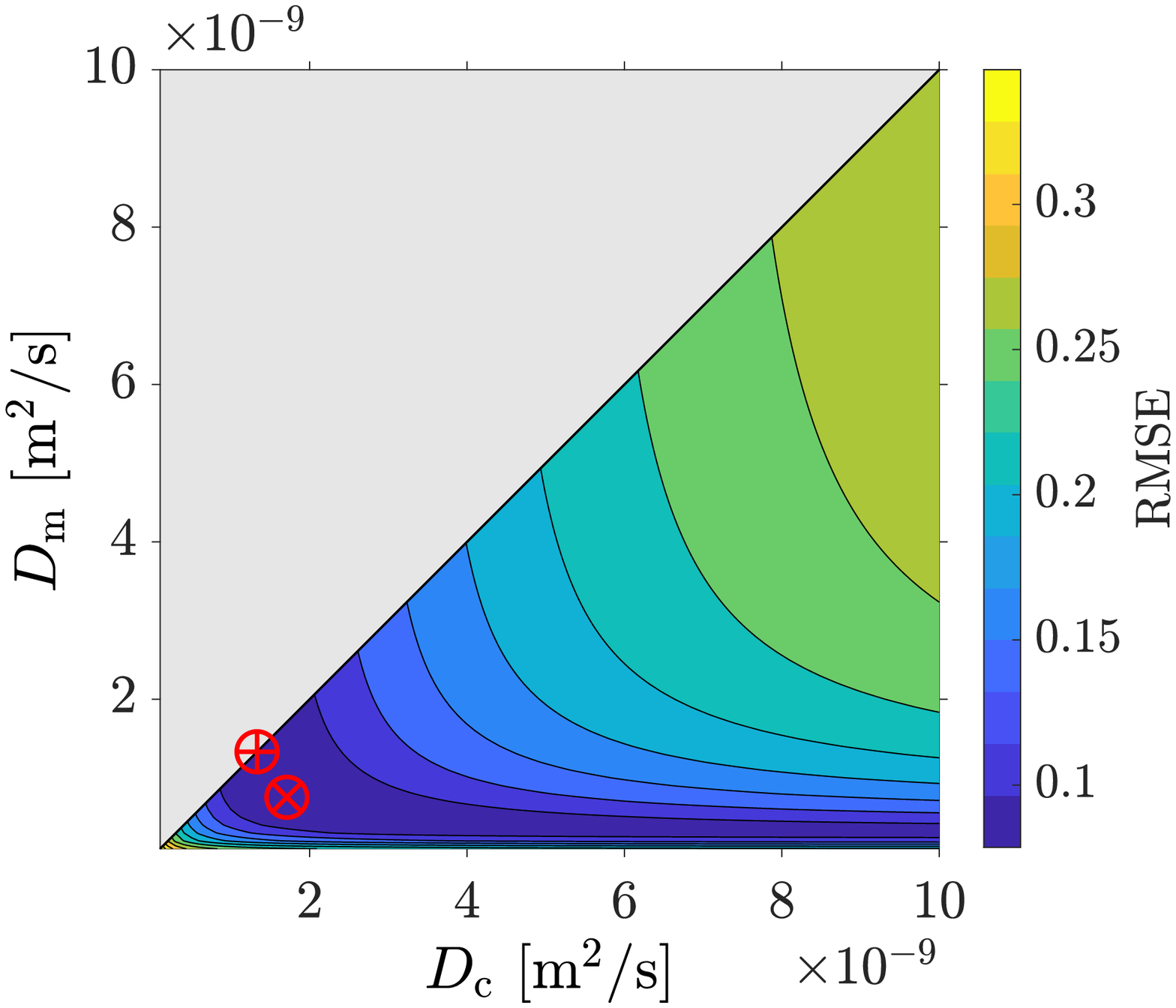}}
    \caption{(a) The release kinetics of glucose from a hydrogel alginate capsule of size $R=1.73\,{\rm mm}$ and membrane thickness $\delta = 50\,\upmu \text{m}$. Square symbols correspond to experimental data taken from Ref.~\cite{Rolland2014}, whereas the solid lines correspond to fits obtained with the present proposed method. (b) The RMSE error between the FD solution and the experimental data while varying the solute diffusion coefficient in both the core and the membrane. The global minimum of RMSE ($\otimes$) gives the diffusion coefficients $D_{\rm c}=17.11 \times 10^{-10}\,{\rm m}^2/{\rm s}$ and $D_\mathrm{m}=7.54 \times 10^{-10}\,{\rm m}^2/{\rm s}$ that fit better the experimental data than the effective diffusion coefficient $D=13.32 \times 10^{-10}\,{\rm m}^2/{\rm s}$ ($\oplus$) obtained when assuming the capsule as a homogeneous sphere.}
    \label{fig:application_Rolland}
\end{figure}

Here, the proposed method is used to determine the membrane permeability to glucose of hydrogel alginate capsules, whose release kinetics has been measured experimentally by Rolland \textit{et al.} \cite{Rolland2014}.
In this study, the size of the capsules is $R=1.73\,{\rm mm}$, and the thickness of the membrane varies within the range $45\, \upmu \text{m} < \delta < 100\,\upmu \text{m}$ that leads to $0.026 < \delta/R < 0.06$.
Only capsules of average membrane thickness $\delta = 50\,\upmu \text{m}$ have been used.
They have been suspended in an aqueous solution whose glucose concentration is kept uniform thanks to stirring.
For this case, the computational domain is limited only to two subdomain layers: the core and the shell excluding the bulk, where the concentration $C_t$ is forced to be uniform but yet varies over time.
The well-stirred solution condition is modeled by Eq.~\eqref{eq:stirring_bc}.
The initial concentration is set to $C_0 = 0.1\,{\rm g}/{\rm mL}$ in the whole capsule, \textit{i.e.} in both the core and the membrane.

The experimental data of Rolland \textit{et al.} are reported in Fig.~\ref{subfig:Rolland_release} as square symbols.
They show the evolution in time of the glucose concentration in the liquid surrounding the capsule.
The fit of these experimental data, while scanning the values of the diffusion coefficients in both the core and the membrane, leads to the RMSE reported in Fig.~\ref{subfig:Rolland_rmse}.
The minimum of the RMSE located on the line $D_{\rm m}=D_{\rm c}$ (symbol $\oplus$) corresponds to the assumption that the capsule is a homogeneous sphere, as done by Rolland \textit{et al.}, whereas the minimum located in the area $D_{\rm m}< D_{\rm c}$ (symbol $\otimes$) considers the composite core-shell nature of the capsule.
The estimated value for a homogeneous sphere is $D = 13.32 \times 10^{-10}\,{\rm m}^2/{\rm s}$, which is close to $D = 14 \times 10^{-10}\,{\rm m}^2/{\rm s}$ measured by Rolland \textit{et al.}
When taking into account the real composite structure of the capsule, the estimated diffusivities are rather $D_{\rm c}=17.11 \times 10^{-10}\,{\rm m}^2/{\rm s}$ and $D_\mathrm{m}=7.54 \times 10^{-10}\,{\rm m}^2/{\rm s}$.
The corresponding release curve fits the experimental data better than the one obtained for the homogeneous sphere, as demonstrated in Fig.~\ref{subfig:Rolland_release}.
Moreover, Rolland \textit{et al.} have pointed out that their measured value is about two times larger than the one in an infinitely dilute solution, that is $6.75 \times 10^{-10}\,{\rm m}^2/{\rm s}$.
They attributed their observed increase in $D$ to the stirring and motion of the capsule that speeds up the release.
Here, the obtained value of $D_{\rm m}$ is close to $6.75 \times 10^{-10}\,{\rm m}^2/{\rm s}$, which is expectedly small.
The flow around the capsule induced by stirring is expected to enhance the release rate due to convection \cite{Kaoui2017,Wang2019,Bielinski2021a}, but not to significantly increase the intrinsic permeability of the capsule membrane.
\begin{figure}
    \centering
    \subfloat[\label{subfig:Koyama_uptake}]{\includegraphics[width=0.45\textwidth]{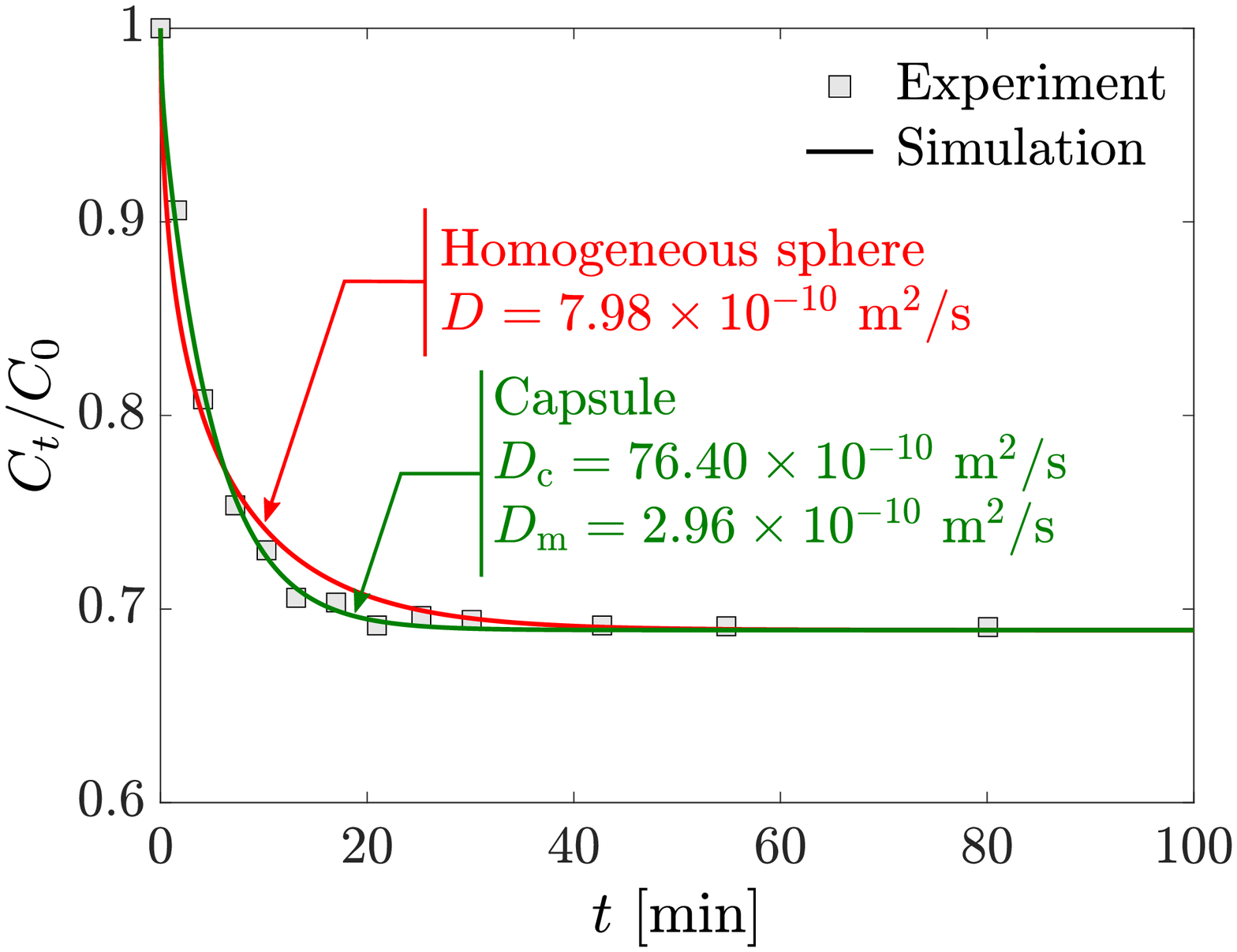}}
    \subfloat[\label{subfig:Koyama_rmse}]{\includegraphics[width=0.45\textwidth]{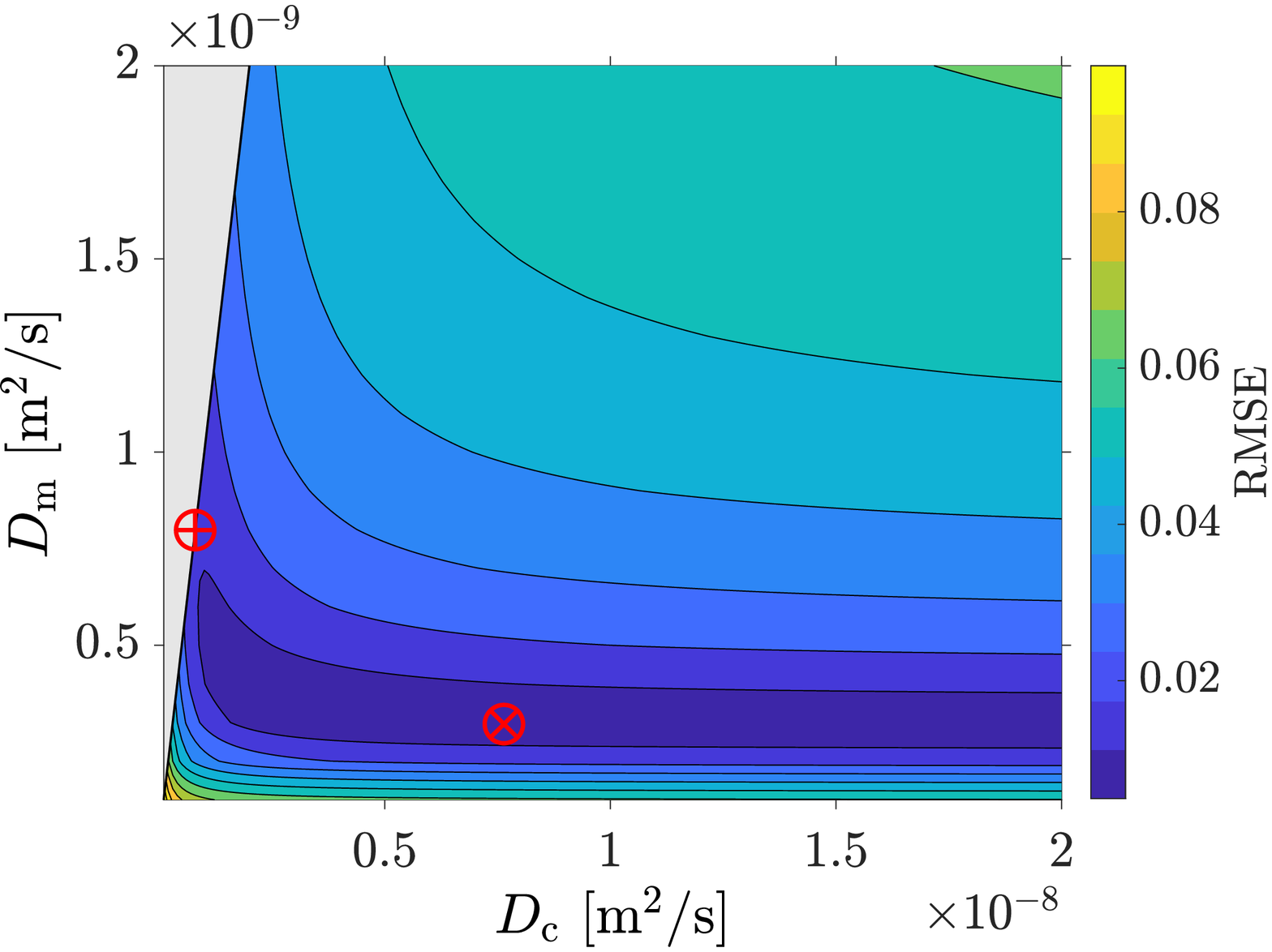}}
    \caption{(a) Evolution of the normalized glucose concentration in the bulk solution over time. Squares: experimental data taken from Ref.~\cite{Koyama2004} showing the glucose absorption kinetics for capsules made of alginate and polyethylene glycol with core radius $R_{\rm c} = 2.17$\,mm and membrane thickness $\delta = 160\,\upmu \text{m}$. Solid line curves: fits obtained while considering the capsule as a homogeneous sphere (red), and with its real core-shell composite structure (green). (b) RMSE between the input experimental data points and the FD solutions obtained over a wide range of the fitting parameters $D_\mathrm{c}$ and $D_\mathrm{m}$. The local minimum along the line $D_\mathrm{m} = D_\mathrm{c}$ (symbol $\oplus$) that corresponds to the case of a homogeneous sphere is obtained at $D= 7.98\times10^{-10}\,{\rm m}^2/{\rm s}$. It is close to the value estimated by Koyama and Seki \cite{Koyama2004} ($D=7.9\times10^{-10}\,{\rm m}^2/{\rm s}$).
    The global minimum is located at $D_\mathrm{c}=76.40\times10^{-10}\,{\rm m}^2/{\rm s}$ and $D_\mathrm{m}=2.96\times10^{-10}\,{\rm m}^2/{\rm s}$ (symbol $\otimes$) that lead to a better fit of the experimental data than the homogeneous sphere.}
    \label{fig:application_Koyama}
\end{figure}
\subsection{Application to experimental data of glucose absorption}
The present characterisation method can also be used in the case of solute absorption.
Here, it is applied to glucose absorption by capsules made of alginate and polyethylene glycol using the experimental data of Koyama and Seki \cite{Koyama2004}.
In this study, the inner core radius of the capsules is $R_\mathrm{c} = 2.17\,{\rm mm}$ and the membrane thickness is $\delta = 160\,\upmu \text{m}$.
The prepared capsules have been immersed into a well-stirred solution of glucose that diffuses to the inner core of the capsules through their membrane.
The authors have measured the variation of the glucose concentration in the surrounding bulk over time, which is reported in Fig.~\ref{subfig:Koyama_uptake} as square symbols.
The RMSE between these experimental data points and the numerical solution obtained by the FD method while varying the fitting parameters $D_\mathrm{c}$ and $D_\mathrm{m}$ is shown in Fig.~\ref{subfig:Koyama_rmse}.
As in the case of solute release, the RMSE is very sensitive to the variations of the membrane diffusivity $D_\mathrm{m}$. 
However, it is slightly affected by changes in the core diffusivity $D_\mathrm{c}$, especially at low membrane diffusivities.
The absorption process is thus mainly controlled by the membrane permeability.
The global minimum (symbol $\otimes$) is obtained at $D_\mathrm{c}=76.40\times10^{-10}\,{\rm m}^2/{\rm s}$ and $D_\mathrm{m}=2.96\times10^{-10}\,{\rm m}^2/{\rm s}$.
The local minimum along the line $D_\mathrm{m} = D_\mathrm{c}$ (symbol $\oplus$) that corresponds to a homogeneous sphere is located at $D_\mathrm{m} = D_\mathrm{c} = D = 7.98\times10^{-10}\,{\rm m}^2/{\rm s}$, and it is very close to $D = 7.9\times10^{-10}\,{\rm m}^2/{\rm s}$ estimated by Koyama and Seki.
The computed absorption curves are plotted in Fig.~\ref{subfig:Koyama_uptake}.
The curve of the capsule model with an internal composite structure again fits perfectly the experimental data.
\section{Conclusions}
\label{sec:conclusions}
The proposed method has been applied to determine the permeability of capsules in the cases of release and absorption of a solute, while considering continuous and well-stirred boundary conditions at the capsule membrane.
It determines the diffusion coefficient in both the membrane and the core, in contrast to the classically used approach that can only give an effective diffusion coefficient for the whole capsule.
The computed diffusion coefficients in the membrane are smaller than the effective diffusion coefficients determined by the classical method.
This is expected since the membrane is a porous medium whose role is to control and slow down the mass transfer.
Furthermore, the present method recovers also the effective diffusion coefficients estimated by the classical method.
Its ease-of-use makes the proposed method an efficient tool for determining accurately the capsule membrane permeability for drug delivery applications.
It can also be used as a computer-aided tool in designing future capsules with desired release or absorption kinetics.
The corresponding Matlab script is provided in Ref.~\cite{Bielinski2021d}.
\section*{Acknowledgements}
The authors thank the Minist\`{e}re de l’Enseignement Sup\'{e}rieur, de la Recherche et de l’Innovation (MESRI) and the Biomechanics and Bioengineering Laboratory (BMBI) for financial support.
\section*{Disclosure statement}
The authors report no conflict of interest.

\end{document}